\documentclass[a4paper,11pt]{article}
\pdfoutput=1 

\usepackage{jcappub} 

\usepackage[T1]{fontenc} 
\usepackage{subfig}

\def\be{\begin{equation}}
\def\ee{\end{equation}}
\def \nn{\nonumber}
\def\bsym{\boldsymbol}
\def\lt{\left}
\def\rt{\right}
\def\vt{\tilde{v}}
\def\Bt{\tilde{B}}
\def \grad{\vec{\nabla}}
\def \vecB{\vec{B}}
\def\barr{\begin{array}}
	\def\earr{\end{array}}
\def\bfig{\begin{figure}}
\def\efig{\end{figure}}

\newcommand{\beq}{\begin{eqnarray}}
\newcommand{\eeq}{\end{eqnarray}}

\title{\boldmath Chiral Battery, scaling laws and magnetic fields }

\author[a]{Sampurn Anand,}
\author[a]{Jitesh R Bhatt,}
\author[a,b]{Arun Kumar Pandey}

\affiliation[a]{Physical Research Laboratory, Ahmedabad, 380009, India}
\affiliation[b]{ Indian Institute of Technology, Gandhinagar, 382424, India}

\emailAdd{sampurn@prl.res.in}
\emailAdd{jeet@prl.res.in}
\emailAdd{arunp@prl.res.in}

\abstract{We study the generation and evolution of magnetic field in the presence of chiral imbalance and gravitational anomaly which gives an additional contribution to the vortical current. The contribution due to gravitational anomaly is proportional to $T^2$ which can generate seed magnetic field irrespective of plasma being chirally charged or neutral. We estimate the order of magnitude of the magnetic field to be $10^{30}$~G at $T\sim 10^9$ GeV,  with a typical length scale of the order of $10^{-18}$ cm, which is much smaller than the Hubble radius at that temperature ($10^{-8}$ cm). Moreover, such a system possess scaling symmetry. We show that the $T^2$ term in the vorticity current along with scaling symmetry leads to more power transfer from lower to higher length scale as compared to only chiral anomaly without scaling symmetry.}

\keywords{Early Universe, Chiral effects, Seed Magnetic field}

\begin{document}
\maketitle
\flushbottom

\section{Introduction}
\label{sec:intro}
In the standard model of cosmology, it is usually assumed that the early
Universe is composed of plasma of elementary particles and the
long range fields are absent. However the observed magnetic field, which
requires initial seed and its subsequent amplification, is one exception.
Moreover, the existence of large scale magnetic field indicates toward its
primordial origin. Till date, various mechanism have been proposed to generate such
magnetic field
\cite{Vilenkin:1991a,Vachaspati_91a, Vilenkin_97,Vachaspati_91b, Enqvist_94,Enqvist_93, Kibble_95, Quashnock_89, Carroll_90, Giovannini_00a, Cornwall_97, Gasperini_95, Lemoine_95, Dolgov_93a, Dolgov_93b, Turner_88}.

Recently, considerable attention is being paid to investigate the role of
quantum chiral anomaly in the generation of primordial magnetic field. It was
argued
that processes in the early Universe, at temperature much above electroweak
scale ($T\sim 100$GeV), can lead to more number of right-handed particles than
left-handed ones and they remain in thermal equilibrium via its
coupling with the hypercharge gauge bosons \cite{Kuzmin:1985mm,tHooft1976az}.
Their number is effectively conserved at those scales which allow us to
introduce the chiral chemical potentials. Furthermore, in presence of either
external gauge field or rotational flow of the fluid, there could be
a current in the direction parallel to the external field or  parallel to the
vorticity generated due to rotational flow of the chiral plasma. The current
along the external magnetic field in presence of chiral imbalance is known as
the ``chiral magnetic current'' and the phenomenon is called the
``Chiral Magnetic Effect'' (CME) \cite{Fukushima:2008xe,PhysRevD.22.3080,Neiman2011,NIELSEN1983389,Alekseev:1998ds}. Similarly, the current parallel to the
rotation axis is known as ``chiral vortical current'' and the phenomenon is
called ``Chiral Vortical Effect'' (CVE) \cite{PhysRevD.20.1807,Johanna:2009,Banerjee2011,son:2009,Landsteiner:2011}. CME and CVE are characterized by the
respective coefficients $\xi^{(B)}$ and $\xi$. 

In the presence
of quantum anomalies for global current there exist a term
$\xi\, \omega^\mu$, with $\omega^\mu \sim \epsilon^{\mu\nu\lambda\rho}\,u_\nu\partial_\lambda u_\rho $, in the conserved current \cite{son:2009}. 
It has been argued that the vorticity induced current $\xi \, \omega^\mu$
is not only allowed, but is required by anomalies  \cite{son:2009}.
By demanding that the derivative of the entropy current $s^\mu$ is non-negative
{\it i.e.} $\partial_\mu\,s^\mu \geqslant 0$, the coefficient $\xi$ can be
expressed in terms of chemical potential, number density, energy
density $\rho$ and pressure density $p$. In \cite{Tashiro:2012mf} authors have
studied the generation and evolution of magnetic field in presence of vortical
current due to chiral imbalance. However, if we have $n_L + n_R = 0$ which
imples $\mu=\mu_L + \mu_R =0$, then the chiral vortical coefficient 
$\xi\sim (\mu_R^2-\mu_L^2)/\sigma$ will vanish. As a consequence, there
will not be any seed magnetic field as discussed in \cite{Tashiro:2012mf}.
Interestingly, there are situations where vortical conductivity is
proportional to $ T^2$ \cite{Landsteiner:2011,Neiman2011}. 
 The origin of $T^2$ term in the vortical conductivity is attributed to the gravitational anomaly which arises when gauge fields are coupled to gravity (for detailed discussion on gravitational anomaly see \cite{AlvarezGaume:1983ig}. A first principle calculation of transport coefficients using Kubo formula leads to anomalous vortical conductivity containing a term $\propto T^2~$ which may be due to  $~C_{\rm g}~\epsilon^{\mu\nu\rho\lambda}~R^\alpha_{~~\beta\mu\nu}~R^\beta_{~~\alpha\rho\lambda}$, with $C_{\rm g}$ being the gravitational anomaly coefficient, in the expression for the anomaly equation \cite{AlvarezGaume:1983ig,Landsteiner:2011}. Although, the gravitational anomaly is fourth order in the metric derivatives though they contributes to first order in the transport coefficients.  One plausible explanation is that the gravitational field in the presence of matter gives rise to a fluid velocity $u^\mu$ e.g. through frame dragging effects and might effectively reduce the number of derivatives that enter in the hydrodynamic expansion \cite{Landsteiner:2011}.

 In the early Universe where chemical potential of
the species is much smaller than the temperature, the dominant
contribution to the vortical current comes from $T^2$.
As a consequence, the
seed magnetic field can be generated in the early Universe irrespective of
fluid being chirally neutral or charged. On the other hand, the equations of chiral
magnetohydrodynamics (ChMHD), in absence of other effects like charge
separation effect and cross helicity, follow unique scaling property and
transfer energy from small to large length scales. 

In this work we will consider the correction to the current density due to
gravitational anomaly and show that the seed magnetic field can be generated
in the every early universe. We will solve the diffusivity equation in an
expanding background. Rather than solving the Navier-Stroke's equation for
velocity spectrum, we will consider the scaling symmetries, consistent with
$T^2$ term in the vorticity current, given in \cite{Yamamoto:2016xtu} to
obtain the velocity spectrum. We show that the amount of energy at large
length scale under scaling symmetry is more than previous study where
only chiral anomaly was considered \cite{Tashiro:2012mf}.\\
 This paper is organized as follows: in section \ref{sec:ch-review} we will
 provide the brief description of the chiral fluid and discuss the generation
 of seed magnetic field due $T^2$ term in the current density in section
 \ref{sec:seed-mag}. We discuss about the scaling properties of the chiral
 plasma in section \ref{sec:scaling}. We will provide the results in
 section \ref{sec:result} and Finally conclude in section \ref{sec:conc}

\section{Overview of Chiral fluid}
\label{sec:ch-review}
%
The dynamics of a  plasma consisting of chiral particles in presence of
background fields is governed by the following hydrodynamic equations
\be
\nabla_\mu\, T^{\mu\nu} = F^{\nu\lambda}\, j_\lambda \, ,
\label{eq:StressCons}
\ee
\be
\nabla_\mu\, j_{\rm v}^\mu  =  0 \, ,
\label{eq:VCons}
\ee
\be
\nabla_\mu\, j_5^\mu  =  C\, E_\mu\,B^\mu \, ,
\label{eq:ACons}
\ee
where the vector current $j_{\rm v}^\mu = j^\mu_R +j^\mu_L$ and chiral current
$j_5^\mu = j^\mu_R -j^\mu_L$ with $j_{R(L)}$ being the current density for 
right (left) handed particles. The chiral anomaly coefficient is denoted by $C$.
In the state of local equilibrium, the energy momentum tensor $T^{\mu\nu}$, the
vector current $j^\mu$ and the chiral current $j^\mu_5$ can be expressed
in terms of the four velocity of the fluid $u^\mu$, energy density $\rho$,
vector charge density $n_{\rm v}$ and axial charge density $n_5$ and these quantities
respectively takes the following form
\be
  T^{\mu\nu}  = (\rho + p)\, u^\mu\, u^\nu\, + p\, g^{\mu\nu}\, ,
  \label{eq:EnergyMomentum}
\ee
\be
  j^\mu_{\rm v} = n_{\rm v}\,u^\mu  + \sigma E^\mu + \xi_{\rm v}\omega^\mu +
  \xi_{\rm v}^{(B)}B^\mu\, ,
\label{eq:VectorCurrent}
\ee
\be
  j_5^\mu = n_5\,u^\mu 
  + \xi_{\rm 5}\omega^\mu + \xi_{5}^{(B)}B^\mu \, ,
  \label{eq:AxialCurrent}
\ee
where $n_{\rm v,5} = n_{R} \pm n_L$,  $\xi_{\rm v,5} = \xi_R \pm \xi_L$, 
  $\xi^{(B)}_{\rm v,5} = \xi^{(B)}_R \pm \xi^{(B)}_L$, 
$\omega^\mu = \epsilon^{\mu\nu\sigma\delta}\,u_\nu \partial_\sigma\,u_\delta$ is the 
vorticity four vector, $E^\mu = F^{\mu\nu}\,u_\nu$, and 
$B^\mu = 1/2\,\epsilon^{\mu\nu\sigma\delta}\,u_\nu\,F_{\sigma\delta}$.
In presence of chiral imbalance and gravitational anomaly, consistency
with second law of thermodynamics ($\partial_\mu\,s^\mu \geq 0$, with $s^\mu$
being the entropy density) demand that the coefficients, for each right and left
particle,
have the following form \cite{son:2009,Landsteiner:2011,Neiman2011}
  \beq
  \xi_i & = & C\,\mu_i^2\,\lt[1-\frac{2\,n_i\,\mu_i}{3\,(\rho + p)}\rt]\, +
  \frac{D\,T^2}{2}\lt[1-\frac{2\,n_i\,\mu_i}{(\rho + p)}\rt] \, ,
  \label{eq:xi} \\
  \xi_i^{(B)} & = & C\,\mu_i\,\lt[1-\frac{n_i\,\mu_i}{2\,(\rho + p)}\rt]\, -
  \frac{D}{2}\lt[\frac{n_i\,T^2}{(\rho + p)}\rt]\, .
  \label{eq:xiB}
  \eeq
The constants $C$ and $D$ are related to those of the chiral anomaly and 
mixed gauge-gravitational anomaly
as $C=\pm 1/4\pi^2$ and $D=\pm1/12$ for right and left handed chiral particles
respectively.
It is now easy to see from eq.(\ref{eq:xi}) and eq.(\ref{eq:xiB}) that for 
$\mu_i/T \ll 1$, which is the scenario in the early universe, 
$\xi_i \sim \frac{D}{2} \, T^2$ and $\xi_i^{(B)} \sim C\, \mu_i$. Note that,
 $\xi_{\rm v} = 0 = \xi_{\rm v}^{(B)}$. 
However, $\xi_5 \sim |D|T^2$ and $\xi_5^{(B)} = |C|\,(\mu_R - \mu_L) = |C|\,\mu_5$. 
We would like to emphasize here that the $\xi_5$ is independent of chiral 
imbalance and depends only on the temperature. Thus, we expect such term to  
dominate at high temperature. We exploit this dominance to generate the magnetic 
field in the early Universe. 

In an expanding background, the line element can be described by 
Friedmann-Robertson-Walker metric,
\be
ds^2 =a^2(\eta)\lt(-d\eta^2 + \delta_{ij}dx^i\,dx^i\rt)\, ,
\ee
where $a$ is the scale factor and $\eta$ is the conformal time. We choose $a(\eta)$ to have dimensions
of length, and $\eta$ , $x^i$ to be dimensionless. Using the fact that the scale factor $a = 1/T$, we
can define the conformal time $\eta = M_*/T$, where $M_* = \sqrt{90/8\,\pi^3\,g_{\rm eff}} M_P$ with
$g_{\rm eff}$ being the effective relativistic degrees of freedom that contributes to the energy density
and $M_P = 1/\sqrt{G}$ being the Planck mass. We also define the following comoving variables 
\be
\vec B_c  =  a^2(\eta)\, \vec B(\eta)\, ,~~~
\mu_c =  a(\eta)\,\mu\, ,~~~\sigma_c = a(\eta)\,\sigma\, ,~~~
T_c  = a(\eta)\,T ,~~~
x_c = x/a(\eta)\, .
\ee
In terms of comoving variables, the evolution equations of fluid and
electromagnetic fields are form
invariant \cite{Holcomb:1989tf,Dettmann:1993zz,Gailis:1995ohk}. Thus, in the
discussion below, we will
work with the above defined comoving quantities and omit the subscript $c$.

Using the effective Lagrangian for the standard model, one can derive the 
generalized Maxwell's equation \cite{Semikoz:2011tm},  
\be
\grad \times \vecB = \vec j\, ,
\label{eq:GenMax}
\ee
with $\vec j = j_{\rm v} + j_5$ being the total current. In the above equation, we 
have ignored the displacement current. Taking $u^\mu = (1,\vec v)$ 
and using eq.(\ref{eq:VectorCurrent})- eq.(\ref{eq:AxialCurrent}), one can show
that
\beq
& j^0 & = n = n_{\rm v} + n_5 \, \nn \\
&\vec{j} & = n\vec v + \sigma(\vec{E} +\vec{v}\times \vec{B})  +
\xi \, \vec{\omega} + \xi^{(B)}\,\vec{B}  \label{eq:totCurrent}\, ,
\eeq
with $\xi = \xi_{\rm v} + \xi_5$ and $\xi^{(B)} = \xi^{(B)}_{\rm v} + \xi^{(B)}_5$. For our analysis, we assume that the velocity field is divergence free,
{\it i.e.} $\grad\cdot\vec v=0$.  
Taking curl of the eq.(\ref{eq:GenMax}) and using the expression for current from
eq.(\ref{eq:totCurrent}), we get
\be
\frac{\partial \vecB}{\partial \eta} =  \frac{n}{\sigma}\,\grad \times \vec v
+\frac{1}{\sigma}\,\nabla^2\vecB + \grad\times(\vec{v}\times \vecB) 
+ \frac{\xi}{\sigma}\,\grad\times\vec\omega
 +  \frac{\xi^{(B)}}{\sigma}\,\grad\times\vecB  \, .
\label{eq:magevol}
\ee
In obtaining the above equation, we have also assumed that chemical
potential  and the temperature are homogeneous.
Note that, in the limit $\sigma\rightarrow \infty$, we will obtain magnetic
fields which are flux frozen. There are other conditions in the cosmological
and astrophysical settings in which the advection term
$\grad\times(\vec{v}\times \vecB)$ can be ignored compared to other terms and
reduces the above equation to linear equation in $v$ and $B$. We ignore
advection term in our analysis.
\section{Chiral battery
}
\label{sec:seed-mag}
In absence of any background magnetic fields $\vecB = 0$, eq.(\ref{eq:magevol}) reduces to 
%
\beq
\frac{\partial \vecB}{\partial \eta}  =
\frac{1}{\sigma}n\grad \times \vec v~ +~
 \frac{1}{\sigma}\,\xi\, \grad\times\vec{\omega}\, .
\label{eq:NoBMaxEq}
\eeq
Note that the two terms on the right hand side of eq.(\ref{eq:NoBMaxEq}) no longer
depends on $B$. This situation is similar to that of the Biermann battery mechanism where 
$\grad \rho\, \times\, \grad p$ term is independent of $B$ \cite{BIERMANN1950}. For Biermann 
mechanism to work, $\grad \rho$ and $\grad p$ have to be in different directions 
which can be achieved if the system has vorticity.
In our scenario, it is interesting to note that even if $n=0$, $T^2$ term in 
$\xi$ will act as a source to generate the seed magnetic field at high temperature.
On the other hand, in presence of finite chiral imbalance (such that $\mu/T \ll 1$) in the early universe, $T^2$ term in $\xi$ will still be the source to generate the seed magnetic field, but non-zero $\xi^{(B)}$ will lead to the instability in the system. An order of magnitude estimate gives the first term of eq.(\ref{eq:NoBMaxEq}) to be 
$ \frac{\alpha}{L}\,\left(\frac{\mu}{T}\right)\,T^2\,v$ while the second term is of the order
$\frac{\alpha}{L^2}\,D\,T\,v$. On comparing the two terms we get a critical length $\lambda_{\rm c} \sim (D/T)(\mu/T)^{-1}$ below which first term will be sub-dominant.
\subsection{Mode decomposition }
In order to solve Eq.(\ref{eq:NoBMaxEq}), we decompose the divergence free vector fields in the polarization modes, $\varepsilon^{\pm}_i$,
\be
\bsym{ \varepsilon}^\pm({\bsym k}) = \frac{\bsym{e}_1({\bsym k})\,\pm \,i\bsym{\,e}_2({\bsym k})}{\sqrt{2}}\, \exp(i\,{\bsym k}\cdot {\bsym x})\, .
\ee
These modes are the divergence-free eigenfunctions of the Laplacian operator
forming an orthonormal triad of unit vectors  $(\bsym{e}_1, \bsym{e}_2, \bsym{e}_3 = \bsym{k}/k)$.
It is evident that $\nabla\cdot\bsym{\varepsilon}^\pm = 0$ and $\nabla\times\bsym{\varepsilon}^\pm = \pm\,k\,\bsym{\varepsilon}^\pm$. For our work $\bsym{\varepsilon}^{\pm*} (-\bsym k) = \bsym{\varepsilon}^\pm (\bsym k)$.
We decompose the velocity of the incompressible fluid as:
\be
\bsym v (\eta, \bsym x) = \int \frac{d^3\,k}{(2\pi)^3}\,
\lt[
  \vt^+(\eta, \bsym k)\bsym{\varepsilon}^{+}(\bsym k) +
  \vt^-(\eta, \bsym k)\bsym{\varepsilon}^-(\bsym k)
  \rt]\, 
\label{eq:Vdecomp}
\ee
where, tilde denote the Fourier transform of the respective quantities. Assuming
no fluid helicity and statistically isotropic correlators,
\beq
\langle \bsym  \vt^{\pm*}(\eta, \bsym k)\, \bsym  \vt^{\pm}(\eta, \bsym q)\rangle & = &(2\,\pi)^3\, \delta^{(3)}(\bsym k - \bsym q)|\bsym v(\eta,k)|^2 \\
\langle \bsym  \vt^{+*}(\eta, \bsym k)\, \bsym  \vt^{-}(\eta, \bsym q)\rangle & = & \langle \bsym  \vt^{-*}(\eta, \bsym k)\, \bsym  \vt^{+}(\eta, \bsym q)\rangle = 0\, ,
\eeq
the kinetic energy density can be given as
\beq
\frac{1}{2}\langle|\bsym v (\eta, \bsym x)|^2\rangle  = 
\frac{1}{2} \int d\log\,k\,E_{\bsym v}(\eta,k) 
 = \int\frac{d^3k}{(2\pi)^3}
 \lt[|\bsym v^+(\eta,\bsym k)|^2\, + \,|\bsym v^-(\eta,\bsym k)|^2\rt] \, .
 \label{eq:ke-spec}
\eeq
Similar to the velocity field, we can decompose the magnetic field as,
\be
\bsym B (\eta, \bsym x) = \int \frac{d^3\,k}{(2\pi)^3}\,
\lt[
  \Bt^+(\eta, \bsym k)\bsym{\varepsilon}^{+}(\bsym k) +
  \Bt^-(\eta, \bsym k)\bsym{\varepsilon}^-(\bsym k)
  \rt]\,.
\ee
The magnetic energy density is then given by
\be
\frac{1}{2}\langle|\bsym B (\eta, \bsym x)|^2\rangle  = 
\frac{1}{2} \int d\log\,k\,E_{\bsym B}(\eta,k)
 = \int\frac{d^3k}{(2\pi)^3}
 \lt[|\bsym B^+(\eta,\bsym k)|^2\, + \,|\bsym B^-(\eta,\bsym k)|^2\rt] \, .
 \label{eq:mag-en-den}
\ee
Mode decomposition of  Eq.(\ref{eq:NoBMaxEq}) gives
\beq
\frac{\partial \Bt^+}{\partial \eta} & = & \frac{1}{\sigma}\lt(n\,k\, + \xi\, k^2\,\rt)\vt^+ \, ,
\label{eq:Bpl}\\
\frac{\partial \Bt^-}{\partial \eta} & = & \frac{1}{\sigma}\lt(-n\,k\, + \xi\, k^2\,\rt)\vt^-
\label{eq:Bmi} \, .
\eeq
In absence of external fields, the energy momentum conservation equation
$\nabla_\mu\, T^{\mu\nu} = 0$ implies that  $n$, $\sigma$ and $\xi$ are
constant over time. Thus, the solution to Eq.(\ref{eq:Bpl}) and Eq.(\ref{eq:Bmi}) is
\be
\Bt^\pm(\eta, \bsym k) =  \frac{1}{\sigma}\lt(\pm
n\,k\, + \xi\, k^2\,\rt)
\int_{\eta_0}^{\eta}\,d\eta'\vt^\pm(\eta', \bsym k)\,.
\label{eq:sol}
\ee
where $\eta_0 = M_*/T_0$ and we have set $T_0 = 10^{10}$ GeV for our work. Multiplying Eq.(\ref{eq:Bpl}) by $\Bt^{+*}$ and Eq.(\ref{eq:Bmi}) by $\Bt^{-*}$, the ensemble average of the combined equation leads to
\beq
\frac{\partial |\Bt^+|^2}{\partial \eta} & = &  \frac{2}{\sigma}\lt(n\,k\, + \xi\, k^2\,\rt)\langle\Bt^{+*}\vt^+\rangle
\label{eq:BplAvg}\\
\frac{\partial |\Bt^-|^2}{\partial \eta} & = &  \frac{2}{\sigma}\lt(-n\,k\, + \xi\, k^2\,\rt)\langle\Bt^{-*}\vt^-\rangle\, .
\label{eq:BmiAvg}
\eeq
On using the solution obtained in Eq.(\ref{eq:sol}) and the velocity decomposition given in Eq.(\ref{eq:Vdecomp}) we obtain
\be
\langle\Bt^{\pm*}(\eta,\bsym k)~\vt^\pm (\eta,\bsym k')\rangle  = 
\,\frac{1}{\sigma}\lt(\pm n\,k\, + \xi\, k^2\,\rt)
\int_{\eta_0}^{\eta}  d\eta '\langle
                      \vt^{\pm*}(\eta',\bsym k)~~\vt^\pm(\eta,\bsym k')
                      \rangle \, .
\label{eq:BVcrossCorrel}
\ee
It is reasonable to expect that any $k$ mode of the fluid velocity  to be
correlated on the time scale $\tau \equiv |\eta -\eta '| \sim 2\pi/k\,v$ and 
uncorrelated over longer time scales, where $v$ represents the average velocity of the fluid within the correlation scale. Thus, we can write
  \be
\langle
\vt^{\pm*} (\eta',\bsym k)~~\vt^\pm(\eta,\bsym k')\rangle  =
\left\{ \begin{array}{ll}
 (2\pi)^3\langle v^\pm(\eta, \bsym k)^2\rangle \delta^{(3)}(\bsym k -\bsym {k'}) & \mbox{ for $ \tau < \frac{2\pi}{k\,v}$}  \\
 \\
 0 & \mbox{ for $\tau > \frac{2\pi}{k\,v}$ } \end{array} \right.
\label{eq:correl-time}
\ee
 On using the above unequal time correlator in eq.(\ref{eq:BVcrossCorrel}), we obtain
 \be
\langle
\Bt^{\pm*}(\eta',\bsym k)\vt^\pm(\eta,\bsym k')\rangle = 
\frac{(2\pi)^3}{\sigma}|v^\pm|^2\lt(\pm n k+ \xi\,k^2\rt)~ f(\eta,k) \delta^{(3)}(\bsym k -\bsym {k'})
\label{correl-time1}
\ee
 where $f(\eta,k) = \eta -\eta_0 $ for $\eta - \eta_0 \leq 2\pi/(k\,v)$ and zero
 otherwise.
 \subsection{Scaling properties in chiral plasma}
 \label{sec:scaling}
 It has been shown that ChMHD equations for electrically neutral plasma with
 chirality imbalance  have
 scaling symmetry under following transformation \cite{Yamamoto:2016xtu},
\begin{align}
\bsym x\rightarrow \ell x\, , \hspace{0.5cm}
\eta\rightarrow \ell^{1-h}\,\eta\, , \hspace{0.5cm}
\bsym v\rightarrow \ell^h\, \bsym  v, \nn \\
\bsym B\rightarrow \ell^{h}\bsym B\,, \hspace{0.5cm} 
\sigma\rightarrow \ell^{1+h}\,\sigma \label{eq:scalingsymm}
\end{align}
with $\ell>0$ is the scaling factor and $h\in \Re$ is the  parameter, under certain conditions \cite{Yamamoto:2016xtu}. With the above scaling, 
the fluid velocity spectrum is given by
\be
E_v(\bsym k,\eta)\,k^{1+2\,h} = \psi_v(\bsym k,\eta)
\label{FluidVelSpec}
\ee
where $\psi_v$ satisfies
\be
\psi(k/\ell, \ell^{1 - h}\, \eta) = \psi(k,\eta)\, .
\label{eq:psi_v}
\ee
On differentiating eq.(\ref{eq:psi_v}) with respect to $\ell$ and taking $\ell = 1$ leads to the following differential equation \cite{Olesen:1996ts}
\be
-k\frac{\partial\psi_v}{\partial k} + (1-h)\, \eta\,\frac{\partial\psi_v}{\partial \eta} =0\, ,
\ee
which can be solved to obtain the form of $\psi$. The solution to the above
differential equation can be given as
\be
\psi_{v}(\eta, k) \propto \,k^m\,\eta^{m/(1-h)}\, ,
\label{ScalePsi}
\ee
where $m$ is a constants and taken as parameter in this work whose value is
fixed by assuming Kolmogorov spectrum for $k>k_i$ and white
noise spectrum for $k<k_i$, where $k_i$ is the wave number in the inertial
range. From
eq.(\ref{FluidVelSpec}) and eq.(\ref{ScalePsi}) , we get
\be
E_v(\eta, k) = v_i^2\,\lt(\frac{k}{k_i(\eta)}\rt)^{-1-2h+m}\,\lt(\frac{\eta}{\eta_0}\rt)^{\frac{m}{1-h}}\,.
\label{eq:ke-spec1}
\ee
where $v_i$ is some arbitrary function which encodes the information about
the boundary conditions.
Therefore, the $k$ and $\eta$ dependence of fluid velocity can be given by
\be
v(k,\eta)  = v_i\,\lt(\frac{k}{k_i(\eta)}\rt)^{(-1-2h+m)/2}\,
\lt(\frac{\eta}{\eta_0}\rt)^{\frac{m}{2(1-h)}}\,.
\label{eq:vel}
\ee
\bfig[!t] 
\subfloat[Subfigure 1 list of figures text][]{
 \includegraphics[width=3.in,height=3.0in,angle=0]{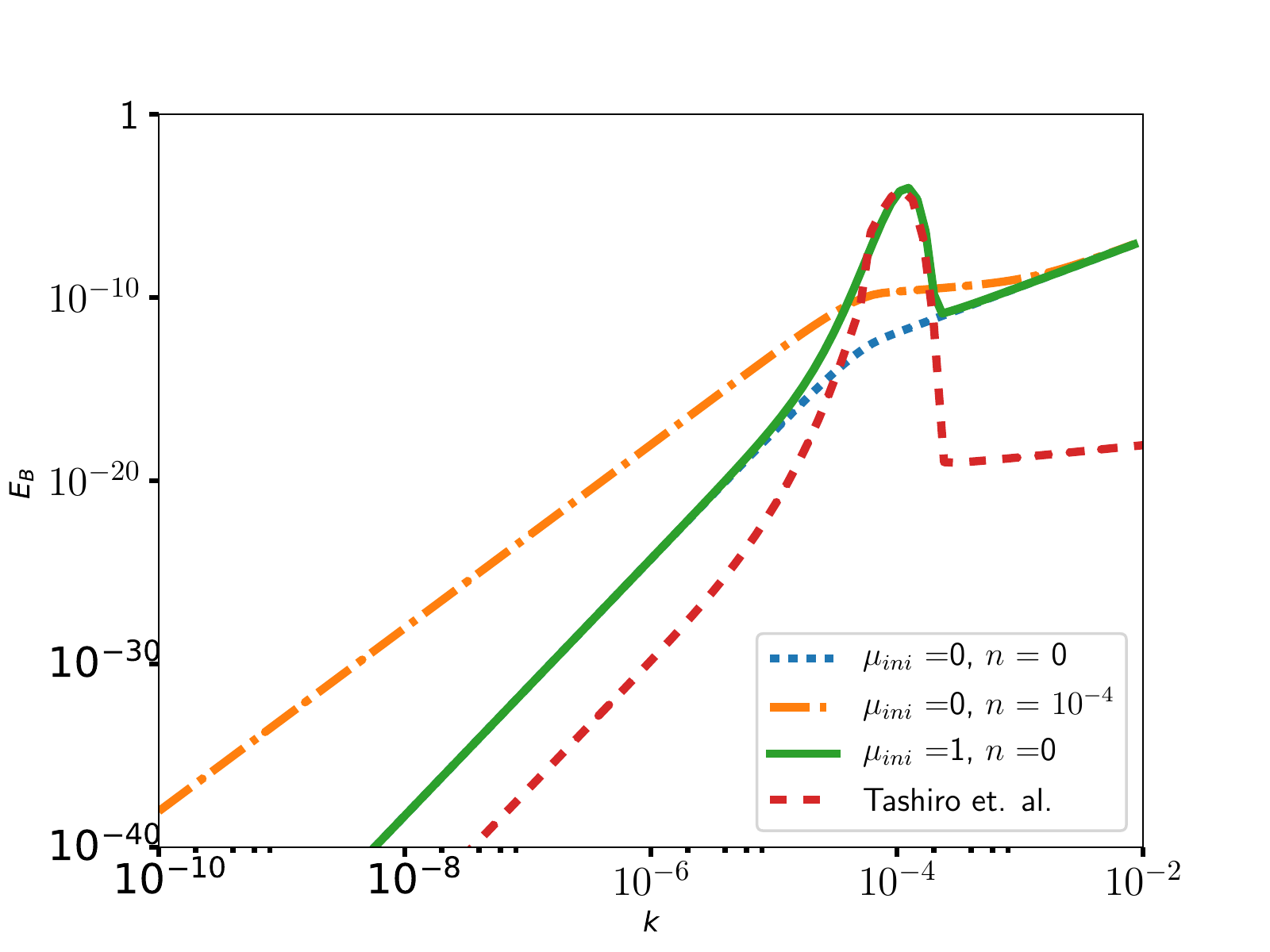}
\label{fig:comp-tashiro}}
\subfloat[Subfigure 2 list of figures text][]{
\includegraphics[width=3.in,height=3.0in,angle=0]{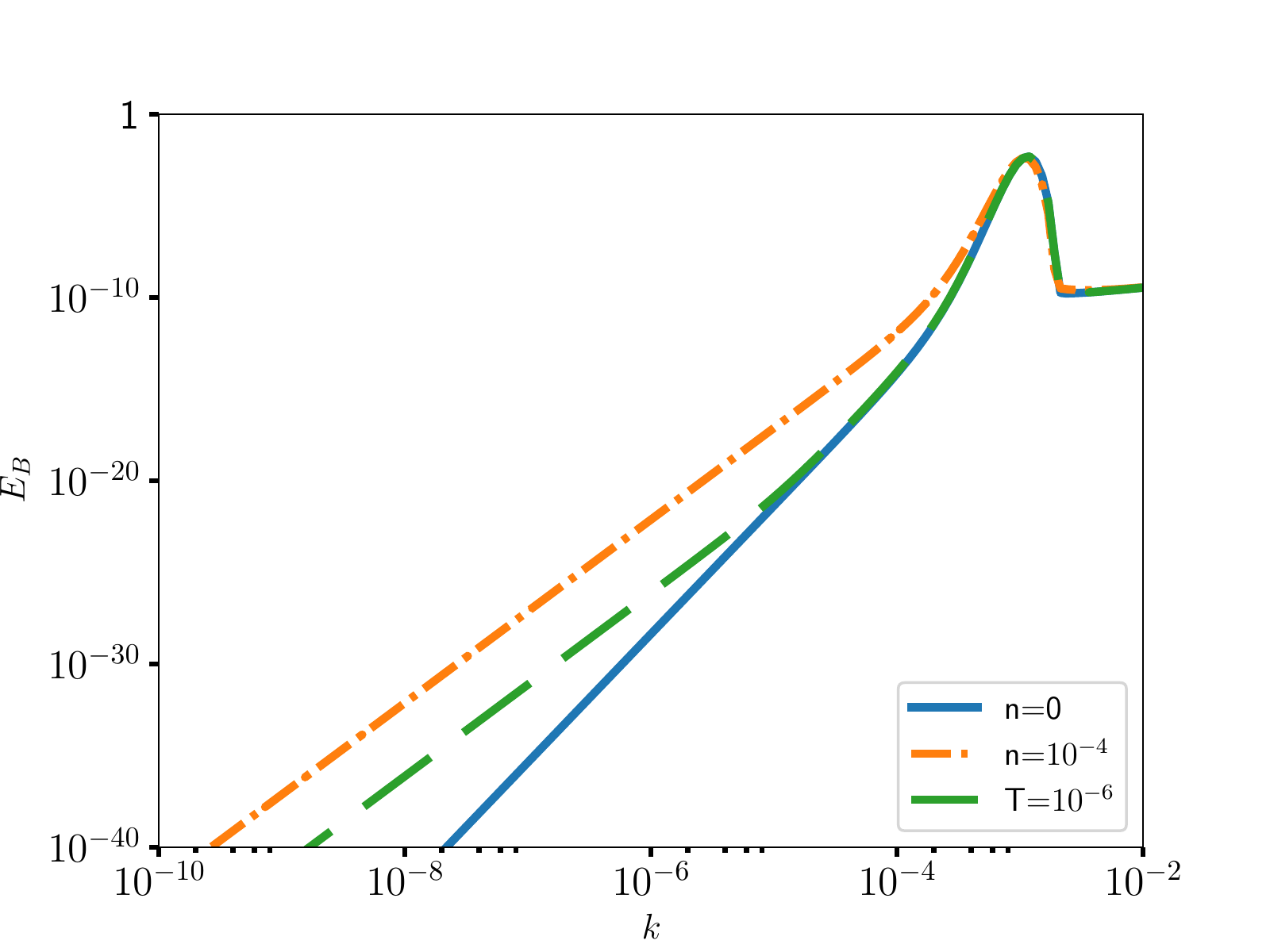}
\label{fig:mag_spec}
	  }
\caption{Magnetic energy spectrum as a function of mode is shown: (a) when
  $n=0$, the $E_B$ goes as $~k^7$ at large length scale. However, the
  power is more than \cite{{Tashiro:2012mf}} at given $k$. (b) when $n \neq 0$,
 $E_B$ goes as $~k^5$ at large length scale. In both the figures we have set $T=10^9$ GeV and $v=10^{-4}$.
}
 \label{fig:Eb_n}
\efig
We are interested in an electrically neutral plasma with chiral charge. For
such a system $h = -1$ \cite{Yamamoto:2016xtu}.
However, in the radiation dominated epoch the fluid velocity does not change for
$k<k_i(\eta)$. Assuming white noise on such scale, which implies $E_v \propto k^3$, we obtain
\be
k_i(\eta) = k_i(\eta_0)\, \lt(\frac{\eta}{\eta_0}\rt)^{1/3}
\ee
Substituting the value of $k_i(\eta)$ in eq.(\ref{eq:vel}) we get
\be
v(k,\eta) = v_i\,\lt(\frac{k}{k_i(\eta_0)}\rt)^{(1+m)/2}\,
\lt(\frac{\eta}{\eta_0}\rt)^{(m-2)/6}\,.
\ee
\bfig[!t] 
\hspace{-0.2cm}
\subfloat[Subfigure 1 list of figures text][]{
 \includegraphics[width=3.1in,height=3.0in,angle=0]{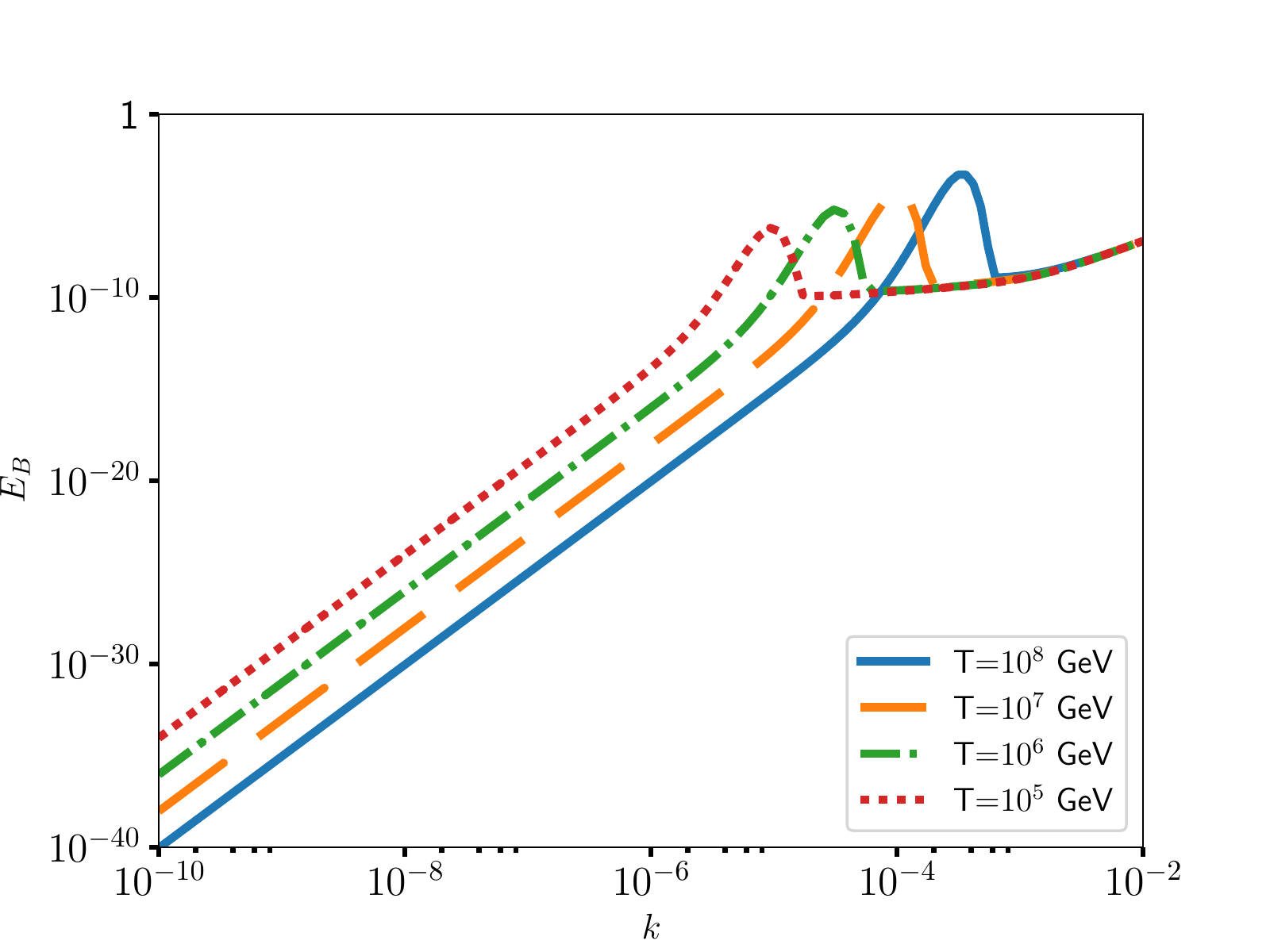}
\label{fig:Eb_T}}
\hspace{-0.3cm}
\subfloat[Subfigure 2 list of figures text][]{
\includegraphics[width=3.1in,height=3.0in,angle=0]{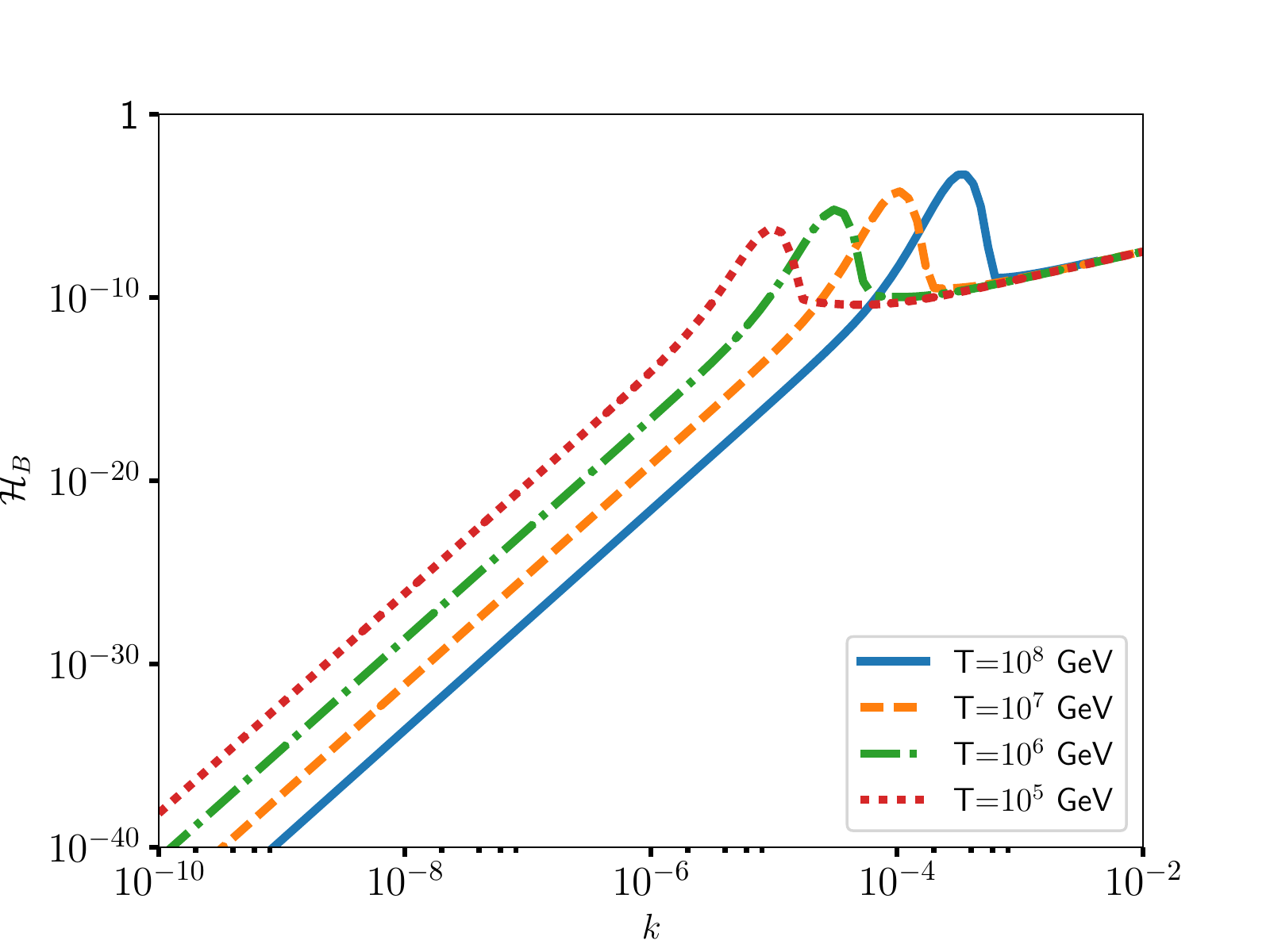}
\label{fig:helicity}
	  }
\caption{(a) Evolution of magnetic energy density $E_B$ as a function of $k$ and (b) Evolution of magnetic 
helicity density $\mathcal{H}_B$  as a function of $k$ for different temperature. 
As temperature ($T\sim 1/\eta$) decreases, peak shifted from higher to lower $k$ values which 
shows transfer of magnetic energy from small to large length scale. In both the figures we have set $v=10^{-4}$
and $n = 10^{-4}.$ 
}
\label{fig:comp_and_hel}
\efig
In absence of kinetic helicity in the fluid, eq.(\ref{eq:ke-spec}) implies
$|v^+|= |v^-| = \pi k^{-3/2}\, v$.
\subsection{Evolution of Magnetic field}
Once seed field is generated it may start influencing the subsequent evolution
of the system at certain mode. In that case, there will be a current flowing parallel to these generated magnetic fields. Therefore, we need to consider the effect of CME in the evolution Eq.(\ref{eq:magevol}). We can decompose this Eq.(\ref{eq:magevol}) in terms of polarization to obtain
\be
  	\frac{\partial |\Bt^\pm|^2}{\partial \eta}  =  \frac{2}{\sigma}\lt(-k^2\pm \xi^{(B)} k\rt) |\Bt^\pm|^2
  	+\frac{2}{\sigma^2}\lt(\pm n\,k\, + \xi\, k^2\,\rt)^2\, |v^\pm|^2\, f(\eta,k)
\label{eq:b-modes}
        \ee
It is evident from the above equation that when the first term dominates over
the second, then
\be
|B^\pm|^2 =|B_0^\pm|^2 \exp\lt(\frac{2\,\eta}{\sigma}\,k_{\rm ins}^2\rt)\,
\exp\lt(\frac{2\,\eta}{\sigma}(k\mp k_{\rm ins})^2\rt) 
\label{eq:modespm}
\ee
where $k_{\rm ins} = \xi^{(B)}/2$.
 From eq.(\ref{eq:b-modes}), evolution of  magnetic energy can be found as
 \beq
 \frac{\partial E_B}{\partial \eta} =\frac{-2\, k^2}{\sigma}E_B+\frac{2\,\xi^{(B)} \, k}{\sigma}\mathcal{H}_B +\frac{2}{\sigma^2}(n^2k^2+\xi^2 k^4)E_v f(\eta,k)
  \label{eq:spectTotalB}
  \eeq
  where $E_B$, $E_v$ and $\mathcal{H}_B$ are respectively magnetic energy, kinetic energy and magnetic helicity which effects the evolution of $\mu_5$ through the following equation
  \be
  \frac{d\mu_5}{d\eta} = -2 \alpha \int \frac{d\ln k}{k} \, \frac{\partial \mathcal{H}_B}{\partial \eta } \, - \Gamma_f\, \mu_5\, ,
\label{eq:chemhel}
  \ee
where $\Gamma_f$ is the chirality flipping rate. The flipping depends on interaction of the right handed particles with the Higgs and their back reaction \cite{CAMPBELL1992118}. At temperature $T\geq 80$ TeV the chirality flipping rate is
much slower than the Hubble expansion rate. Therefore, eq.(\ref{eq:chemhel}) can be rewritten as
\be
\frac{\partial}{\partial\eta}\lt[
\mu_5 + 2\alpha\int \frac{d\ln k}{k}{\mathcal H}_B
\rt] = 0\, .
\ee
The terms in the bracket is constant in time. Therefore, the helicity will be produced at the cost of chiral 
imbalance. However, when temperature drops to an extent that the flipping rate becomes comparable to the 
expansion rate of the Universe, maximally helical field will be generated. In order to get the complete picture
of generation and evolution of the magnetic field,
we solve eq.(\ref{eq:b-modes}) along with eq.(\ref{eq:chemhel}).
\section{Results}
\label{sec:result}
In the early Universe, much before electroweak symmetry breaking, collision of
bubble walls during first order phase transition at  GUT scale, can generate 
initial vorticity in the plasma. With these assumptions, we
have shown that magnetic field can be generated due to any of the two terms on 
the right hand side of eq.(\ref{eq:NoBMaxEq}).
It is also important to note that, second
term have purely temperature dependence ($\propto D\, T^2\, \vec{\omega}$)
which comes due to gravitational anomaly.
In absence of any background field, the two term $n\,k$ and $\xi\,k^2$
in the right hand side of eq.(\ref{eq:b-modes}) can be source of the seed field.
Further, when $n=0$, only $\xi^2\,k^4 $ term acts as a source to
generate the seed magnetic field. However,  there may not be any instability in the system 
if $\xi^{(B)} = 0$ (Fig(\ref{fig:comp-tashiro})). It is also evident form 
eq.(\ref{eq:b-modes}) and eq.(\ref{eq:mag-en-den}) that the magnetic energy 
spectrum will go as
$\sim k^7$ at large length scale (Fig(\ref{fig:comp-tashiro})). This
feature is similar to \cite{Tashiro:2012mf}. However, for a given $k$ the power
transferred to larger length scale is more with the scaling symmetry
(see Fig(\ref{fig:comp-tashiro})). On the other hand, when $n \neq 0$, then
$n\,k$ term dominates over $\xi \, k^2$ term at larger length scale. 
Consequently, the energy spectrum goes as $\sim k^5$ 
(see Fig.(\ref{fig:mag_spec}). We estimate the strength of magnetic field produced to be: $\frac{B}{T^2}  \approx  \frac{\alpha}{D} \left(\frac{\mu}{T}\right)^2 T L(\eta)$, where $L(\eta)=v\eta\approx \frac{D}{T}\left(\frac{\mu}{T}\right)^{-1}$. For temperature $T=10^9$~ GeV and $\mu/T\approx 10^{-6}$, the strength of the magnetic field is of the order of $10^{30}$ G at a length scale of the order of $10^{-18}$ cm, which is much smaller than the Hubble length ($10^{-9}$ cm) at that temperature. 

It is evident from eq.(\ref{eq:b-modes}) that once the magnetic field of
sufficiently large strength is produced, it starts influencing the subsequent
evolution of the system and the magnetic energy grow rapidly.
This phenomenon is known as the chiral plasma instability
\cite{Akamatsu2013, Akamatsu:2014yy}. In this regime, we solve the
eq.(\ref{eq:b-modes}) and obtain mode expression in (\ref{eq:modespm}), which
clearly shows that the modes of the magnetic fields grows exponentially  for
the wave number $k\simeq k_{ins}/2\approx C\mu_5/2$. As mentioned earlier,
for $T\geqslant 80$ TeV is slower than the expansion rate of the Universe, 
one can safely ignore $\Gamma_f\,\mu_5$ term in eq.(\ref{eq:chemhel}). However, 
the magnetic helicity generated due to $DT^2$ will drive the evolution of 
chemical potential. Below $80$ TeV, flipping rate becomes comparable to the expansion 
rate of the Universe and the chirality of the right particles changes to the left handed.
So to know the complete dynamics of the magnetic field energy, we have solved 
the coupled equations (\ref{eq:b-modes}) and (\ref{eq:chemhel}) simultaneously. 
In Fig(\ref{fig:Eb_T}) we have shown the variation of magnetic energy spectrum
$E_B$ with $k$ for different temperature.
Since, the chemical potential is a decreasing function of time, so it is
expected that with decrease in temperature the instability peak will shift
towards the smaller $k$ which implies the transfer of energy from small to
large length scale. Similarly, we have shown the magnetic helicity spectrum in
Fig(\ref{fig:helicity}).
\section{Conclusion}
\label{sec:conc}
In the present work, we have discussed the generation of magnetic field due
to gravitational anomaly which induces a term $\propto T^2$ in the vorticity
current. The salient feature of this seed magnetic field is that it will
be produced at high temperature irrespective of the whether fluid is charged
or neutral. 
In ref. \cite{Yamamoto:2016xtu}, authors have shown that the equations of 
chiral magnetohydrodynamics (ChMHD), in absence of other effects like charge 
separation effect and cross helicity, follow
unique scaling property and transfer energy from small to large length scales
known as inverse cascade.
Under this scaling symmetry more power is transferred from lower to
higher length scale as compared to only chiral anomaly without 
scaling symmetry.
\bibliographystyle{JHEP} 
\bibliography{ref} 

\end{document}